\documentstyle[preprint,aps,epsf,epsfig]{revtex}
\newcommand{\bff}{\rm}
\newcommand{\AmS}{{\protect\the\textfont2  A\kern-.1667em\lower.5ex
\hbox{M}\kern-.125emS}}
\begin{document}
\title{Spin Wave Response in the Dilute Quasi-one 
Dimensional Ising-like Antiferromagnet 
CsCo$_{0.83}$Mg$_{0.17}$Br$_3$} 
\author{Y. S. Yang\thanks{Email:{\tt ysyang@hyowon.pusan.ac.kr}}}  
\address{Research Center for Dielectric and Advanced Matter Physics, 
Department of Physics in Graduate School, 
Pusan National University, Pusan, Korea 609-735} 
\author{F. Marsiglio, M. Madsen} 
\address{Department of Physics, University of Alberta, 
Edmonton, Alberta T6G2J1, Canada}
\author{B. D. Gaulin}
\address{Department of Physics and Astronomy, McMaster University, 
Hamilton, Ontario L8S 4M1, Canada}
\author{R. B. Rogge}
\address{Neutron Program for Materials Research, Chalk River Laboratories, 
Chalk River, Ontario, K0J 
1P0, Canada}
\author{J. A. Fernandez-Baca }
\address{Solid State Division, 
Oak Ridge National Laboratory, Oak Ridge, Tennessee 37831-6393}
\date{\today}
\maketitle 
\begin{abstract} Inelastic neutron scattering profiles of 
spin waves in the dilute quasi-one-dimensional Ising-like antiferromagnet 
CsCo$_{0.83}$Mg$_{0.17}$Br$_3$ have been investigated. Calculations 
of S$^{xx}(Q,\omega)$, based on an effective spin Hamiltonian, 
accurately describe the experimental spin wave spectrum of the 2$J$ mode. 
The $Q$ dependence of the energy of this spin wave mode follows the 
analytical prediction 
$\omega_{xx}(Q)$=(2J)(1-5$\epsilon^{2}cos^{2}Qa+2\epsilon^{2}$)$^{1/2}$,
calculated by Ishimura and Shiba using perturbation theory.
\end{abstract}
The quantum nature of spin-1/2 one-dimensional 
magnetic systems manifests itself in a number of interesting properties 
that have been studied both experimentally and theoretically 
for many years \cite{mattis93}. The spin wave excitation spectrum around 
the pure Ising energy consists of a continuum of states, accompanied 
by the propagation of domain wall pairs of the soliton response. 
The existence of a band of spin wave states and the soliton 
response have been confirmed within the quasi-one-dimensional system, 
CsCoM$_3$(M=Br, Cl), by Satija {\it et al.} \cite{satija80} 
and Nagler {\it et al.} \cite{nagler82}. 

The spin Hamiltonian describing the dynamics of 1-D Ising-like 
antiferromagnets, including the exchange mixing produced 
by inter-chain and intra-chain interactions, was determined 
to be \cite{nagler83}
$$
H = 2J\sum[S_j^zS_{j+1}^z+\epsilon(S_j^xS_{j+1}^x+S_j^yS_{j+1}^y)
]+h\sum(-1)^jS_i^z. 
\eqno(1)
$$
Here $J$ is the exchange coupling between nearest 
neighbour Co ions in a chain, $\epsilon$ is the parameter that 
distinguishes the Ising limit ($\epsilon = 0$) from the 
Heisenberg limit ($\epsilon = 1$), $h$ is an effective (staggered) 
field, which arises from exchange mixing and coupling between chains. 

The Hamiltonian (1) may also model the case where a non-magnetic 
ion (Mg) is substituted for the magnetic ion (Co), which gives a 
distribution of finite length spin chains. {\bff Finite length 
chains are of interest because they allow one to study surface modes
with neutron scattering. Moreover, comparison with theory is more rigorous
since the full spectrum, $S^{xx}(Q,\omega)$, can be computed exactly for
Hamiltonian (1) in small chains.}
Nagler {\it et al.} \cite{nagler84} 
measured the spin wave spectrum of such a dilute one-dimensional chain, 
CsCo$_{0.83}$Mg$_{0.17}$Cl$_3$, with inelastic neutron scattering and 
carried out a detailed comparison with theoretical results based on 
such a Hamiltonian. Their results confirmed the necessity of the staggered 
field.  Nonetheless, the calculated spin wave spectrum at the zone center 
was in disagreement with experiment, and the relative intensities between the 
surface mode ($\omega\sim J$) and the bulk mode ($\omega\sim 2J$) were not 
in quantitative agreement. Furthermore, only two wave vectors were examined, 
therefore up until the present, a full experimental study of 
the spin wave dispersion has been lacking. Such a dispersion 
relation was calculated by Ishimura and Shiba (IS) \cite{is743}, 
and it remains an unchallenged prediction relevant to the effective Hamiltonian 
described above. 

In this {\bff paper, by measuring the spin wave response for several $Q$-vectors
throughout the Brillouin zone}, it is shown that a refined 
calculation of S$^{xx}(Q,\omega)$ quantitatively  
describes the 
spin wave spectrum associated with both the $J$ and the $2J$ mode, and 
that the $Q$-dependence 
of the spin wave energy of this mode indeed follows the IS prediction:
$$
\omega_{xx}(Q) = (2J)(1-5\epsilon^{2}cos^{2}Qa+2\epsilon^{2})^{1/2}.
\eqno(2)
$$

CsCoBr$_3$ belongs to the 
space group $D_{6h}^{4}$ and displays room temperature lattice 
constants a=7.529 $\AA$ and c=6.324 $\AA$. The material consists of 
stacked triangular lattices composed of chains of Co$^{++}$ ions 
arranged parallel to the c-axis. The Co$^{++}$ ions on a given 
chain interact strongly with one another, with exchange 
constant $J$. Long range, three-dimensional ordering occurs at $T_N=28.4$ K, below
which CsCoBr$_3$ enters a partially-paramagnetic, 3-sublattice Neel state.  On further cooling, 
an additional transition occurs, associated with the ordering of the paramagnetic site.  
With the addition of non-magnetic impurities, Mg (17$\%$), in 
place of Co, the ordered states are severely altered in such a way
that order parameter measurements show upwards curvature to temperatures 
as low as 3 K\cite{rogge93}. {\bff In our measurements we stay well above these
temperatures, with the intent of avoiding three dimensional and ordering effects}. 

CsCo$_{1-x}$Mg$_{x}$Br$_3$ is a good candidate system for the 
investigation of dilution effects as the properties of the pure system 
have been well determined.  In addition, for neutron studies,  
the bromide is expected to be preferable to the chloride
as the incoherent scattering cross section of Br is considerably less 
than Cl (0.1 barns compared with 5.1).  {\bff In addition the absorption cross-section
is considerably lower for Br than for Cl}. The dilute system allows for study 
of both the $2J$ mode, which involves a flip of a spin
in the bulk of a finite chain as well as for the 
$J$ mode which arises from spin flips
that occur at either end of the finite chains.

Inelastic neutron scattering experiments 
were carried out on the HB2 triple axis spectrometer at the HFIR research 
reactor of ORNL. Measurements were made with fixed final neutron energy 3.52 THz, 
using Si(111) as the monochromator and pyrolytic (002) as the analyzer. 
Collimation of 40' was chosen for both the incident and 
scattered beams.  The resulting energy resolution was 
about 0.25 THz. A pyrolytic graphite filter was used in 
the scattered beam to suppress the higher order contamination. 

For the dilute system, the 
transverse spin wave response $S^{xx}(Q, \omega)$can be written in 
terms of the response function for a single chain of 
length $N$, $S_N^{xx}(Q, \omega)$:
$$
S^{xx}(Q, \omega) = \sum\limits_N u^{N} (1-u)^2 S_N^{xx}(Q, \omega),
\eqno(3)
$$
where the single chain response is
$$
S_N^{xx}(Q, \omega)=\sum\limits_E|<E|S_N^{xx}(Q)|G>|^2\delta(\omega-E).
\eqno(4)
$$
Here the spin-flip operator is given by
$$
S_N^{xx}(Q, \omega)=\sum\limits_{j=1}^N e^{iQj}{\bf S}_j^{x},
\eqno(5)
$$
where the summation of $j$ is over the length of 
the chain, and ${\bf S}_j^x$ is the usual $x$-component of 
the spin operator for the spin at the $j$th site.
In these expressions, we have used $u$ to denote 
the concentration of magnetic ions.  The summation over the various 
chain lengths is therefore subject to the constraint that the total 
number of magnetic ions in the entire chain equals this concentration.
We use the ket $|E>$ to denote an eigenstate with energy $E$ 
and $|G>$ is the ground state with energy taken to be zero. 

The distribution of chain lengths is assumed to be based on
random substitution of non-magnetic Mg ions for Co in the chain.
This is the same percolation assumption used by Nagler 
{\it et al.} \cite{nagler84} for CsCo$_{0.83}$Mg$_{0.17}$Cl$_3$.

The total $S^{xx}(Q,\omega)$ for the system is then
obtained by first adding the response of chains 
with $N \le 11$, weighted according to the above-mentioned 
percolation theory. The resulting 
response is then scaled from $N = 11$ to larger sizes by 
multiplying the computed response by two factors, $F_{2J}^{\infty}/F_{2J}^{11}$ for 
the $2J$ band and $F_{J}^{\infty}/F_J^{11}$ for the $J$ band. 
The scaling functions 
are 
$$
F_{2J}^M =\sum\limits_{N=2}^M (N-2)(1-u)^2u^N,\, \, \,
F_{J}^M =\sum\limits_{N=2}^M 2(1-u)^2u^N,
\eqno(6)
$$
{\bff 
and are given by \cite{nagler84}
$$
F_{2J}^M = x^3 - x^{M+1}[(M-2)(1-x) + 1]
\eqno(7a)
$$
and
$$
F_{J}^M = 2x^2 (1 - x)(1-x^{M-1}).
\eqno(7b)
$$ 
The F-functions with $M = \infty$ are those that would be obtained
from an Ising system (at $T = 0$), with an average number of chains of
length $N$ equal to $x^N(1-x)^2$ (according to percolation theory \cite{nagler84}).
The calculated responses up to some $M$ ($M = 11$ in our case) are
weighted by these factors in order to account for the presence of larger
chains (which is non-negligible). }

The spin wave response at $13.3$ K in CsCo$_{0.83}$Mg$_{0.17}$Br$_3$, is shown in Fig. 1, 
for a variety of wave vectors spanning the one-dimensional Brillouin zone
from the zone boundary (0, 0, 2.5) to the zone centre (0, 0, 3).  
All the constant {\bf Q} scans taken were of the form (0, 0, L) 
and sensitive only to transverse spin correlations, hence
$S^{xx}(Q, \omega)$, due to the sensitivity of the neutron scattering cross section to spin
components normal to {\bf Q}.  The experimental and calculated peak intensities 
have been scaled to agree at a single wavevector, (0, 0, 2.5).
The data shows two clearly defined 
modes with energy $J$ and $2J$, which are identified 
as magnetic in origin, both by their fall off at large wave vector 
due to the magnetic form factor and by the temperature dependence of 
this scattering.  The phonon peaks, which were also found for the pure 
sample \cite{nagler83}, are indicated by the 
letter P. 

The solid curves are calculations 
of $S^{xx}(Q,\omega)$ weighted, as described above, 
for chains with $ N \le 11$.  We checked the appropriate scaling relations by 
calculating the spectrum for $N \le 14$, and utilizing 
the same procedure.  The results changed very little; indeed, the 
maximum spin chain length originally utilized 
by Nagler {\it et al.} \cite{nagler84} ($N \le 8$) also gave a fairly 
accurate representation of the full distribution.  This result is 
not surprising as the average chain length is 5.
These spectra were calculated at $T = 13.3$ K, with 
parameter values similar to those used to describe
the pure sample, $J = 1.55 THz$, 
and $\epsilon = 0.18J$ \cite{nagler82}. The staggered 
field parameter $h = 0.05J$ and a convolution of the resolution 
function of the instrument with the expected Lorentzian 
function{\cite{villain80}} with full width at half 
maximum $\Delta=$0.1$J$ for the spin wave peak were used without 
further adjustable parameters.

Typically the value of the exchange constant $J$ is determined by  
the peak position of the $2J$ mode at the  
zone boundary.  However, as the peak positions of the two modes depend on  
both $h$ and $\epsilon$ as well, we have fit the spectrum for 
$J$, $h$ and $\epsilon$ simultaneously to give optimal agreement at 
all wave vectors studied.  
A slight discontinuity occurs in  
the calculated curves in Fig. 1 at an energy between 
the $J$ and $2J$ modes (1.5$J$).  This artifact arises because of  
the different scaling employed for each frequency range, as    
discussed above. As can be seen from Fig. 1 the description of the
experiment by theory is quantitatively very good at all wavevectors
across the 1D Brillouin zone.

IS calculated the dispersion for the $2J$ mode, using a combination of
perturbation theory and the previously known second moment 
result{\cite{is743}}, for the transverse spin wave response 
in the absense of a staggered field term.  Their resulting prediction is 
Eq. (2). We have evaluated the energies corresponding to the center of mass of 
the calculated spectra for all wave vectors in Fig. 1 and have 
found that the $\omega$-$Q$ dispersion is indeed well described by IS theory.
In the figure, the energies corresponding to the center of mass are slightly 
larger than the absolute values calculated from Eq. (2).  However,
this is largely due to the influence of the staggered 
field which raises the energy of the spectra by about 0.13$J$ compared 
with the $h$=0 case at the zone boundary. 
It is worth emphasizing that the excellent description
of the scattering line shape by the calculated structure factor  
implies the IS energy dispersion for the $2J$ mode.
These results clearly indicate that the effective  
Hamiltonian (1) describes very accurately the dispersion of
the $2J$ mode. 

The dispersion relation obtained from the fits in Fig. 1 is shown in Fig. 2.
The open circles (2$J$ mode) are the energies of the center of mass of
the spectra. The filled circles show the 2$J$ mode energies corrected 
for the effect of finite staggered field so that they can be compared directly
with Eq. (2).  The solid line represents the 
IS theoretical expression, Eq. (2).  In this figure, the
squares show the dispersion of $J$ mode using the 
same procedure as was used for 2$J$ mode energies.

In Fig. 3, we show a two dimensional contour map 
of the calculated inelastic spectrum
as a function of energy and wave vector.
In this figure, the parameters used are the same as those 
employed in the fits to the data shown in Fig. 1. 
This scattering shows the spin wave continuum as 
predicted by IS, giving the antisymmetric bow-tie shape for the 2$J$ mode.
The spin wave continuum broadens appreciably  
as the zone center is approached, to the values $\sim$ $\pm 4\epsilon J$, consistent
with previous theortical work{\cite{is743,dCG66}}. 

Fig. 4 shows a two dimensional contour map
of the calculated spin wave spectrum
as a function of staggered field, for the zone center wavevector.
The pronounced increase in the ratio of intensities between the 
2$J$ and $J$ modes, as well as the movement of spectra towards higher energy 
with increasing $h$ can be clearly seen.
Both the description of the ratio of the relative intensities in the 2$J$ and $J$
modes in the experimental spin wave scattering intensity, as well as the description 
of the asymmetry of the spin wave lineshapes have been improved by adjusting $h$ and $\epsilon$.

There has been some debate regarding the use of a next nearest neighbour 
exchange interaction (NNN) for the 1-D system in the pure 
sample {\cite{matsu,goff}}.
We have investigated this effect and found the NNN term is not required \cite{goff}
for a satisfactory description of the data. 
We have seen that the staggered field plays a key role in obtaining 
the very good correspondence between the experimental and the calculated lineshapes.

To conclude, {\bff by measuring the spin wave response 
for a variety of scattering wavevectors,}
we have shown that a quantitative understanding of the transverse
spin wave spectrum $S^{xx}(Q,\omega)$ in the quasi-one-dimensional, 
diluted spin-1/2  chains system CsCo$_{0.83}$Mg$_{0.17}$Br$_3$ can be obtained
through an effective spin Hamiltonian (1).  This result provides a stringent
test of the IS perturbative form for the dispersion relation; {\bff by attempting
fits without the staggered field term (and failing), we have highlighted the
importance of such a term, as first suggested by Nagler {\it et al.} \cite{nagler83}.}

This work was supported by the Electron Spin Science Center 
founded by the Korea Science and Engineering Foundation, by
the Natural Sciences and Engineering Research Council of Canada, and
by the Canadian Institute for Advanced Research.
Some of the calculations were performed on the multi-node SGI
parallel processor at the University of Alberta.


\vskip10pc]
 
\begin{figure}[h]
\epsfig{figure=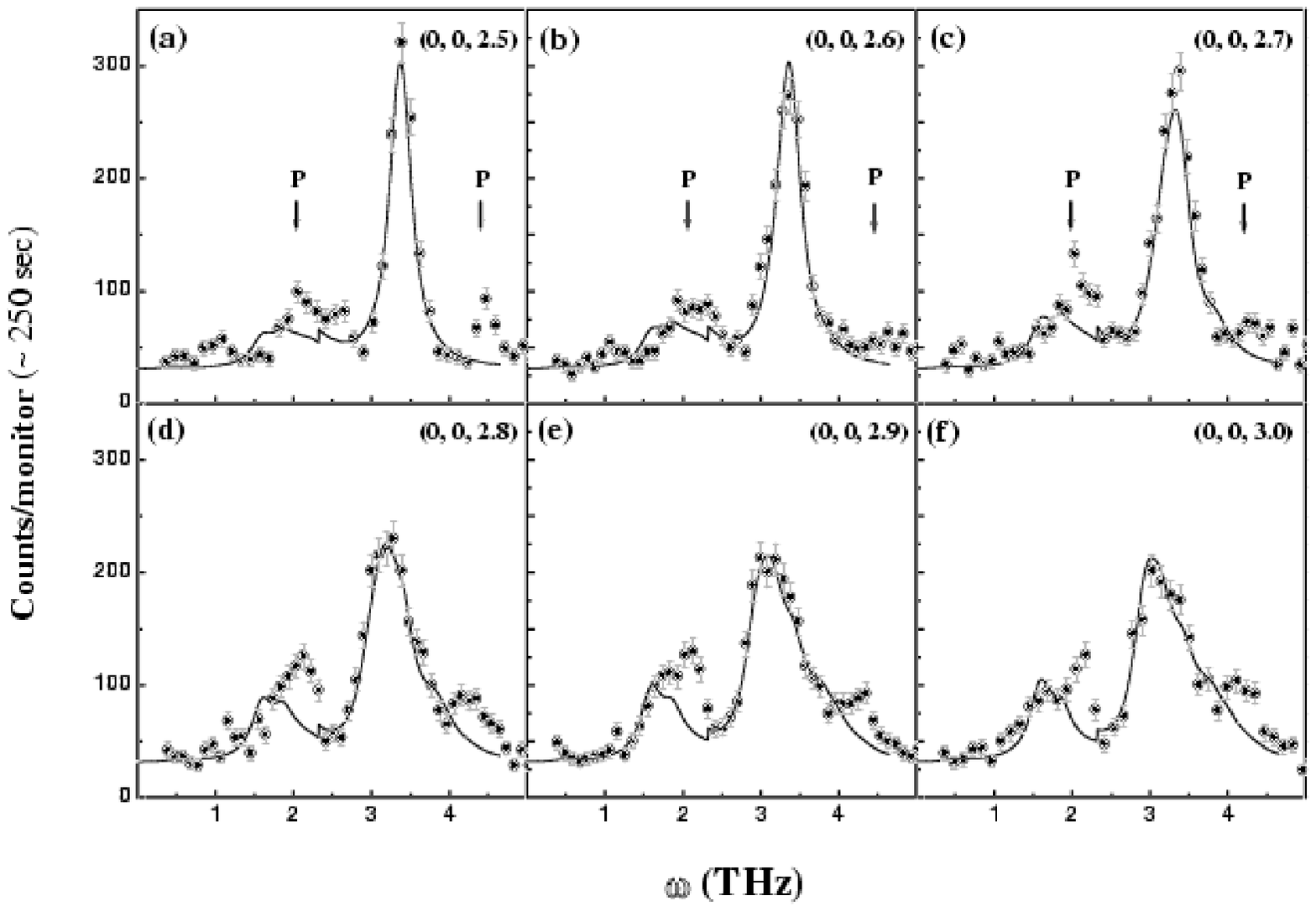,height=14cm,width=16cm}
\caption{Inelastic  neutron scattering at several wave vectors of
the form (0, 0, L),
at 13.3 K, from CsCo$_{0.83}$Mg$_{0.17}$Br$_3$ are shown.
The solid curves are a high quality fit to the data due to
theoretical calculations of $S^{xx}(Q,\omega)$
given by Eq. (3). The parameters used in the calculations are
$J$=1.55 THz, $h$=0.05$J$, $\epsilon$=0.18$J$, $\Delta$=0.1$J$, T=13.3 K.}
\label{fig1}
\end{figure}

\begin{figure}[h]
\epsfig{figure=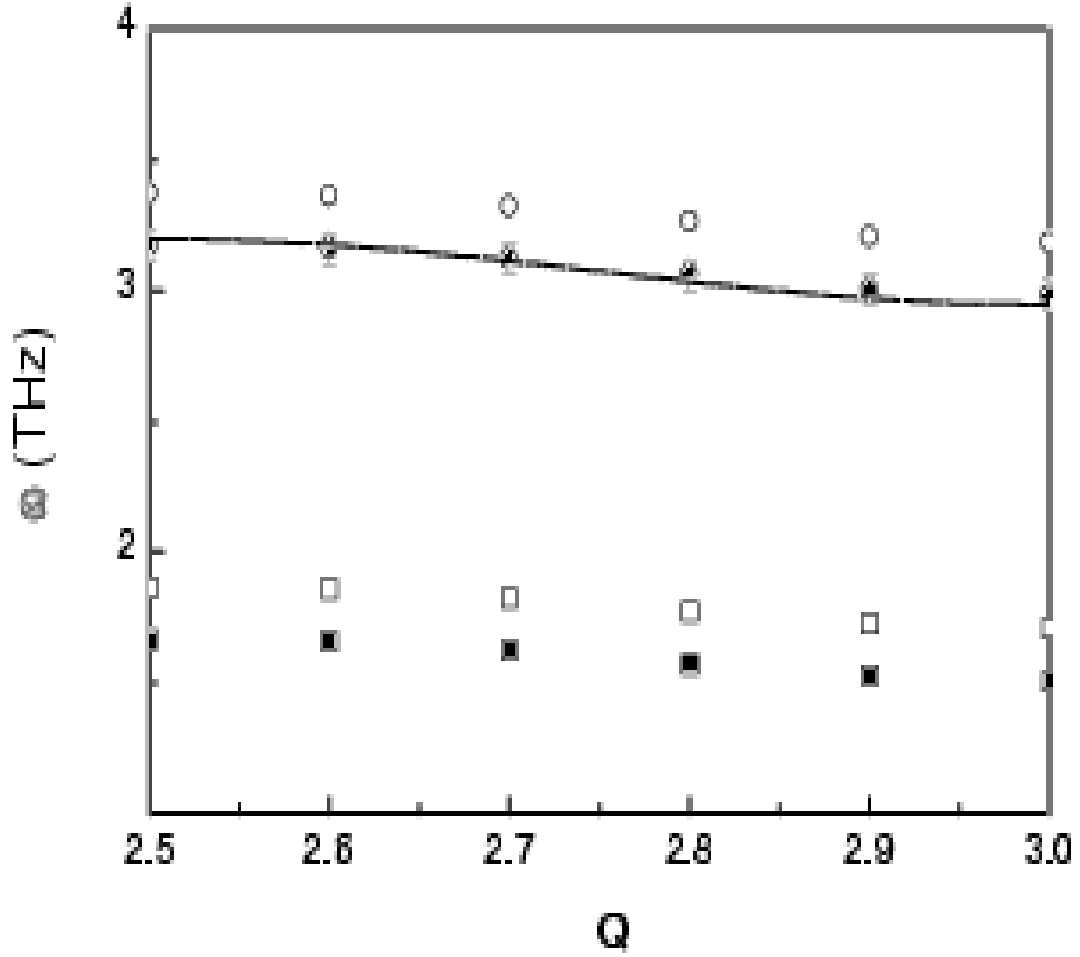,height=14cm,width=16cm}
\caption{The dispersion relation obtained from Fig. 1 for the 
2$J$ ($J$) mode is shown as the open circles (squares).
The filled circles are these energies with the value 
0.13$J$ (0.2 THz) subtracted from open ones. The line corresponds
to the IS theoretical expression, Eq. (2).}
\label{fig2}
\end{figure}

\begin{figure}[h]
\epsfig{figure=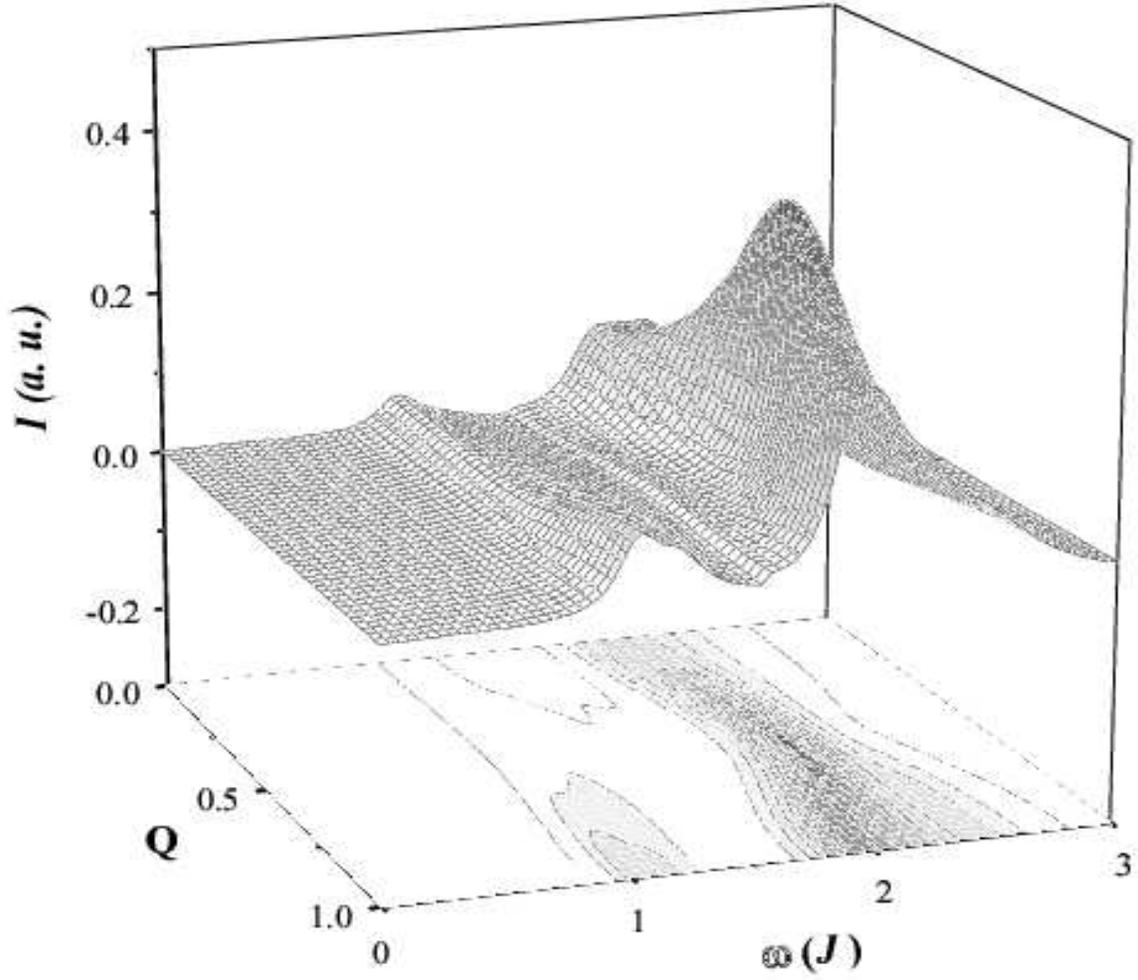,height=14cm,width=16cm}
\caption{The calculated two dimensional contour map of 
the spin wave spectra $S^{xx}(Q,\omega)$.  The parameters used
in the calculation 
are $J$=1.55 THz, $h$=0.05$J$, $\epsilon$=0.18$J$, T=13.3 K 
and $\Delta$=0.1$J$.}
\label{fig3}
\end{figure} 

\begin{figure}[h]
\epsfig{figure=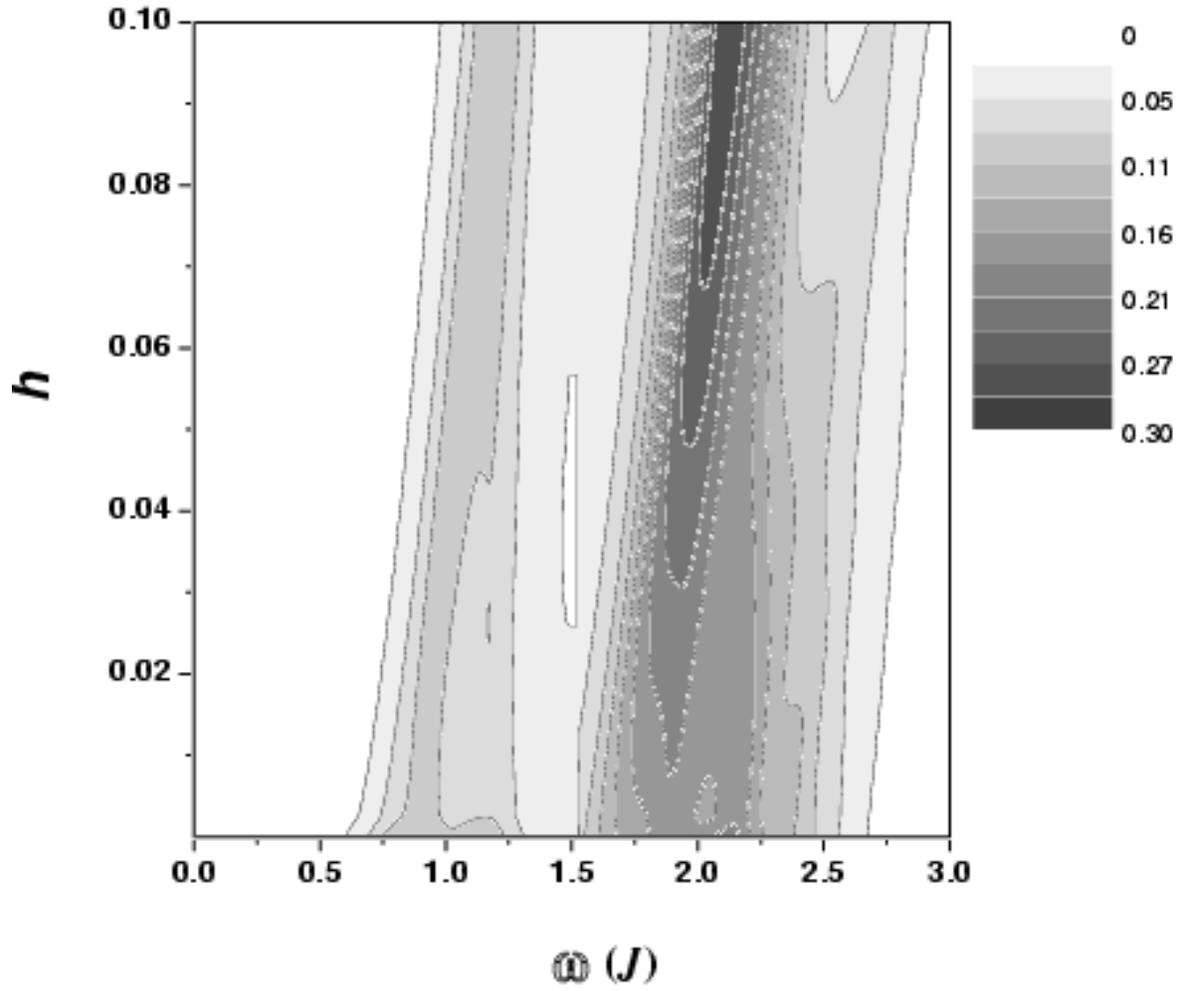,height=14cm,width=16cm}
\caption{The calculated two dimensional contour map of the 
spin wave spectra are shown as a function of
$\omega$ and the staggered term $h$ for the zone center wavevctor. 
The parameters in the calculation are those used in Fig. 3.}
\label{fig4}
\end{figure}


\end{document}